\documentclass[12pt]{article}
\usepackage{a4wide}
\usepackage{amssymb}
\usepackage{graphicx}
\begin{document}
\setcounter{table}{0}

\mbox{}
\vspace{2truecm}
\linespread{1.1}

%%%%%%%%%%%%%%%%%
\centerline{\LARGE \bf On the 5d instanton index as a Hilbert series}

\vspace{2truecm}

\centerline{
    {\large \bf Diego Rodr\'{\i}guez-G\'omez${}^{a}$} \footnote{d.rodriguez.gomez@uniovi.es}
    {\bf and}
    {\large \bf Gabi Zafrir${}^{b}$} \footnote{gabizaf@techunix.technion.ac.il}}

\vspace{1cm}
\centerline{{\it ${}^a$ Department of Physics, Universidad de Oviedo}} \centerline{{\it Avda. Calvo Sotelo 18, 33007, Oviedo, Spain }}
\vspace{1cm}
\centerline{{\it ${}^b$ Department of Physics, Technion, Israel Institute of Technology}} \centerline{{\it Haifa, 32000, Israel}}
\vspace{1cm}

\centerline{\bf ABSTRACT}
\vspace{1truecm}

The superconformal index for $\mathcal{N}=2$ 5d theories contains a non-perturbative part arising from 5d instantonic operators which coincides with the Nekrasov instanton partition function. In this note, for pure gauge theories, we elaborate on the relation between such instanton index and the Hilbert series of the instanton moduli space. We propose a non-trivial identification of fugacities allowing the computation of the instanton index through the Hilbert series. We show the agreement of our proposal with existing results in the literature, as well as use it to compute the exact index for a pure $U(1)$ gauge theory.

\noindent

\newpage

\section{Introduction}

Supersymmetric gauge theories in 5d automatically come with conserved topological global currents of the form $j=\star\,{\rm Tr}\,F\wedge F$  \cite{Seiberg:1996bd}. The electrically charged particles are instantons --which in 5d are particle-like excitations--. Their role is crucial in the dynamics of 5d gauge theories leading, under some circumstances, to enhanced global symmetries on the Higgs branch  \cite{Seiberg:1996bd}. Very recently, strong evidence in favour of this non-perturbative enhancement due to instantonic particles has been given in \cite{Kim:2012gu} (see also \cite{Iqbal:2012xm}) through the computation of the exact superconformal index, which admits an expansion in characters of the enhanced global symmetry group.

In order to compute the index, one considers the Euclidean theory in radial quantization and chooses a supercharge and its complex conjugate. Primary operators annihilated by this subalgebra contribute to the index weighted by their representation under all other commuting charges. In 5d the bosonic part of the $\mathcal{N}=1$ superconformal algebra is $SO(2,\,5)\times SU(2)_R$, where $SU(2)_R$ is the R-symmetry. In turn $SO(2,\,5)$ contains the dilatation operator as well as a compact $SO(5)_L$ acting on the $S^4$. The maximal compact subgroup is $[SU(2)_1\times SU(2)_2]_L\times SU(2)_R$. Calling the $U(1)$ Cartans respectively $j_1,\,j_2,\,R$, the generators commuting with the chosen supercharge are $j_2$ and $j_1+R$. Then, the index reads \cite{Bhattacharya:2008zy, Kim:2012gu}

\begin{equation}
\mathcal{I}={\rm Tr}\,(-1)^F\,e^{-\beta\,\Delta}\,x^{2\,(j_1+R)}\,y^{2\,j_2}\,\mathfrak{q}^{\mathfrak{Q}}\,,\qquad \Delta=\epsilon_0-2\,j_1-3\,R\,,
\end{equation}
where $\mathfrak{Q}$ collectively stands for global symmetries --including the instanton current-- with associated fugacities collectively denoted by $\mathfrak{q}$. As the index does not depend on $\beta$, only states whose scaling dimension satisfies $\epsilon_0=2\,j_1+3\,R$ contribute. In \cite{Kim:2012gu} it was shown that the index admits a path integral representation leading to

\begin{equation}
\label{indexstructure}
\mathcal{I}=\int \mathcal{D}\alpha\,\mathcal{I}_{{\rm pert}}\,\mathcal{I}_{{\rm inst}}\,,
\end{equation}
where $\int \mathcal{D}\alpha$ stands for the integration over the gauge group with the suitable Haar measure, while $\mathcal{I}_{\rm pert}$ and $\mathcal{I}_{\rm inst}$ stand respectively for the perturbative and instantonic contributions to the index.

The perturbative contribution is easily computed by gluing the appropriate building blocks \cite{Kim:2012gu}: one first needs to construct the single-particle index adding one factor of $i_V$ for each vector multiplet and one factor of $i_M$ for each half-hypermultiplet. The corresponding functions read

\begin{equation}
i_V=-\frac{x\,(y+y^{-1})}{(1-x\,y)\,(1-x\,y^{-1})}\,\chi_{\rm Adj}\, ,\qquad i_M=\frac{x}{(1-x\,y)\,(1-x\,y^{-1})}\,\chi_{\rm M}\, ,
\end{equation}
where $\chi_{{\rm Adj}}$ and $\chi_{\rm M}$ stand for the characters of the representation under the gauge group and other possible global non-instantonic symmetries. Note that as far as $\mathcal{I}_{\rm pert}$ (and $\mathcal{I}_{\rm inst}$) is concerned, the gauge fugacities $\alpha$ appear as global symmetries --it is the $\int \mathcal{D}\alpha$ in eq. (\ref{indexstructure}) what projects to gauge-invariants--. Thus, we will loosely refer to all non-Lorentz fugacities as global symmetries. Upon taking the plethystic exponential of the single-particle index one immediately finds $\mathcal{I}_{\rm pert}$. \footnote{Recall that the plethystic exponential is defined as ${\rm PE}[f(\vec{x})]=e^{\sum_{n=1}^{\infty}\,\frac{f(\mathbf{x}^n)}{n}}$, where $\mathbf{x}$ stands for the set of all fugacities on which the single-particle index $f$ might depend on. Besides $\mathbf{x}^m=(x_1^{m},\,x_2^m,\,\cdots)$. }

The instantonic contribution is, in turn, much harder to compute. On general grounds it localizes on instantonic configurations around both north (anti-instantons) and south (instantons) poles of the $S^4$  \cite{Kim:2012gu}, so that locally one needs to compute the path integral on $S^1\times \mathbb{R}^4$ over the solution space of the constraint $F^+=0$ for the south pole and $F^-=0$ for the north pole. This is precisely the Nekrasov instanton partition function, which can be thought of as the Witten index of a supersymmetric quantum mechanics on the moduli space of instantons $\mathcal{M}$. Making use of the results in \cite{Kim:2011mv}, in  \cite{Kim:2012gu} the instanton partition function for $USp(2)$ theories with one antisymmetric hyper and $N_f$ fundamental hypers were computed and the emergence of the global $E_{N_f+1}$ was shown. Alternatively, the same result was recovered in \cite{Iqbal:2012xm} by using the topological vertex.

The Nekrasov instanton partition function for pure instantons, regarded as the Witten index of a supersymmetric quantum mechanics, can be thought of as the equivariant index $\sum_i(-1)^i\,{\rm Tr}_{H^{(0,\,i)}}\,g$ where $g$ is an element of the global symmetry group (see \cite{Tachikawa, Nekrasov:2004vw} for very comprehensive introductions). As such, it is intimately related with the Poincare polinomial a.k.a. Hilbert series  \cite{Benvenuti:2010pq,Hanany:2012dm} of the corresponding instanton moduli space $\mathcal{M}$, a generating functional of the (graded) coordinate ring $\mathbb{C}[\mathcal{M}] =\bigoplus_i \mathcal{H}^{(0,\,i)}$ --where the elements of $\mathcal{H}^{(0,\,i)}$ are holomorphic degree $i$ homogeneous polynomials-- whose unrefined version is defined as $\sum_i\,{\rm dim}\,\mathcal{H}^{(0,\,i)}\,t^i$. In a sense, this object counts BPS wavefunctions on $\mathcal{M}$, so it is natural to expect it to be related to the Witten index of the quantum mechanics on $\mathcal{M}$. While this connection has been implicitly suggested in the literature (see \textit{e.g.} \cite{Kim:2011mv,Keller:2011ek}), its status is somewhat vague. In this note we make it precise for the class of 5d pure gauge theories and check it in a few simple cases showing explicit agreement of the instanton index arising from the Hilbert series with known results in the literature.

The structure of this note is as follows: in section \ref{motivation} we make explicit our conjecture for the  computation of the instanton index as a Hilbert series. In fact, equation (\ref{conjecture}) contains our main result. In section \ref{BO} we test our conjecture in the particular example of the $USp(2)$ theory whose global symmetry has been conjectured to be enhanced non-perturbatively to $SU(2)$ \cite{Seiberg:1996bd} and find explicit agreement with \cite{Kim:2012gu, Iqbal:2012xm}. In section \ref{U(1)} we compute the exact index of a pure $U(1)$ gauge theory. We conclude in section \ref{conclusions} with some remarks.

\section{Hilbert series and instanton index}\label{motivation}

We consider a 5d pure gauge theory with gauge group $G$. As reviewed above, the index contains a contribution from instantonic operators. Such contribution factorizes into the product of $G$-instantons localized around the south pole and $G$-anti-instantons localized around the north pole of the $S^4$ \cite{Kim:2012gu}. Let us denote the instanton partition function for instantons around the south pole by $\mathcal{I}_{{\rm inst}}^{\rm S}(q)$. Denoting by $q$ the instanton current fugacity, such function can be expanded as

\begin{equation}
\mathcal{I}_{\rm inst}^{\rm S}=\sum_{k=0}^{\infty}\mathcal{I}_{{\rm inst}}^{\rm (k)}\,q^k\, ,
\end{equation}
so that $\mathcal{I}_{\rm inst}^{\rm (k)}$ is the $k$-instanton partition function (of course, $\mathcal{I}_{\rm inst}^{(0)}=1$). As such, it depends on the Lorentz fugacities $x,\,y$ as well as on the G-fugacities $\alpha_i$.  Recall that, from the point of view of the instanton index, gauge symmetries look like global symmetries, as it is $\int \mathcal{D}\alpha$ in eq.(\ref{indexstructure}) what projects to gauge-singlets. Thus, as far as $\mathcal{I}_{\rm inst}$ is concerned, we can regard $G$ as a global symmetry.\footnote{As we will be interested in pure gauge theories, for $G=SU(N)$ we can have as well global baryonic symmetries. These can be thought of as the $U(1)$ part in $U(N)$. In those cases the global symmetry of $\mathcal{I}_{\rm inst}$ is the full $U(N)$ to which we will also refer as $G$. In fact, consistently, in the ADHM construction, the flavor symmetry of the dual ADHM quiver is $G=U(N)$.}

On the other hand, the instanton index for anti-instantons localized around the north pole can be easily obtained  \cite{Kim:2012gu} as $\mathcal{I}_{\rm inst}^{\rm N}(q)=\mathcal{I}_{\rm inst}^{\rm S}(q^{-1})$. Then, the whole instanton contribution to the index is just $\mathcal{I}_{\rm inst}=\mathcal{I}_{\rm inst}^S\,\mathcal{I}_{\rm inst}^{\rm N}$. It is then clear that the quantities of interest are the $G$ $k$-instanton partition functions $\mathcal{I}_{\rm inst}^{\rm (k)}$. Our claim is that these are just close relatives of the Hilbert series of the $G$ $k$-instanton moduli space.

Following \cite{Benvenuti:2010pq,Hanany:2012dm}, the $G$ $k$-instanton on $\mathbb{C}^2$ Hilbert series can be constructed by considering the auxiliary gauge theory --sometimes called the \textit{Kronheimer-Nakajima quiver}-- whose Higgs branch realizes the desired instanton moduli space through the ADHM construction. As it is well-known, the $k$-instanton moduli space on $\mathbb{C}^2$ is realized as the Higgs branch of a gauge theory with gauge group $\widehat{G}_{\rm k}$, an adjoint hypermultiplet and $N$ fundamental hypermultiplets transforming under a global $G$ symmetry. The gauge group $\widehat{G}_{\rm k}$ is the ADHM dual gauge group (see for example \cite{Nekrasov:2004vw} for an explicit description in various cases).

The auxiliary dual ADHM theory will generically have an $SU(2)$ global symmetry associated to the adjoint hypermultiplet in addition to a global $G$ symmetry associated to the flavor symmetry. Thus, the Hilbert series on the Higgs branch computed following the techniques in \cite{Benvenuti:2010pq,Hanany:2012dm} will depend on a fugacity $\hat{y}$ for the $SU(2)$ and on $\hat{\alpha}_i$ fugacities associated to $G$. Besides, it will depend on a fugacity $\hat{x}$ standing for the dimension of the operators --actually the fugacity $t$ corresponding to the degree of the grading as defined above-- which can be thought of as an $\mathbb{R}$ fugacity. As such, indeed only positive powers of $\hat{x}$ appear in the Hilbert series. Thus, the Hilbert series on the Higgs branch of the ADHM auxiliary theory can be written as $HS_{\rm k}=HS_{\rm k}(\hat{x},\,\hat{y},\,\hat{\alpha}_i)$.

In turn, the Nekrasov partition function $\mathcal{I}_{\rm inst}^{\rm (k)}$ depends on an element $g$ of a compact global symmetry group involving both the Lorentz fugacities $\{x,\,y\}$ as well as as the global symmetry fugacities $\alpha_i$ for $G$. For the latter, it is clear that the $G$ fugacities $\alpha_i$ in $\mathcal{I}_{\rm inst}^{\rm (k)}$ will be identified with the $G$ fugacities $\hat{\alpha}_i$ in $HS_{\rm k}=HS_{\rm k}(\hat{x},\,\hat{y},\,\hat{\alpha}_i)$. It remains to clarify the mapping between $(x,\,y)$ and $(\hat{x},\,\hat{y})$.

To that matter, recall that both $\hat{x}$ and $\hat{y}$ have a clear geometrical meaning. Indeed, consider the case of pointlike instantons on $\mathbb{C}^2$. As these have no internal structure, the Hilbert series will be purely geometrical. Furthermore, it can be written \cite{Benvenuti:2006qr, Benvenuti:2010pq} as ${\rm PE}[\hat{x}\,(\hat{y}+\hat{y}^{-1})]$, which shows that the moduli space is constructed with two dimension 1 generators transforming as an $SU(2)$ doublet. On the other hand, upon introducing complex coordinates $\{z_1,\,z_2\}$, $\mathbb{C}^2$ is invariant under $SU(2)_a\times SU(2)_b$, acting each on the doublets $\{z_1,\,z_2\}$ and $\{z_1,\,\bar{z}_2\}$. Then the coordinate ring on $\mathbb{C}^2$ is constructed in terms of monomials of the generic form $z_1^m\,z_2^n$ which obviously have definite transformation properties under $SU(2)_a$. Hence the $SU(2)$ with fugacity $\hat{y}$ associated to the adjoints directly maps to the $SU(2)_a$ geometric symmetry on $\mathbb{C}^2$. Besides, the degree of the monomial, basically given by $\delta=n+m$, directly maps to the fugacity $\hat{x}$. As in polar coordinates, both $z_{1,\,2}$ are proportional to the radial coordinate, so we have that $\hat{x}$ is an $\mathbb{R}$ fugacity.

In turn, both $(x,\,y)$ in $\mathcal{I}_{\rm inst}^{\rm (k)}$ are $SU(2)$ fugacities corresponding to the (compact) global symmetry group element $g$. Note that $SU(2)$ characters $[n]_z$ \footnote{We use the notation $[n]_z$ for $SU(2)$ characters, where $[1]_z=z+z^{-1}$, $[2]_z=z^2+1+z^{-2}$ and so on.} are invariant under $z\leftrightarrow z^{-1}$. This ``symmetry" is inherited by the generating function. In fact, because of the same reason, it is easy to check that $HS_{\rm k}(\hat{y})=HS_{\rm k}(\hat{y}^{-1})$. Obviously, since $\hat{x}$ is not an $SU(2)$ fugacity, $HS_{\rm k}(\hat{x})\ne HS_k(\hat{x}^{-1})$. However, it is possible to construct an invariant quantity under this transformation by considering $\widehat{HS}_k(\hat{x})=\hat{x}^{a}\,HS_{\rm k}(x)$ such that $\widehat{HS}_{\rm k}(\hat{x})=\widehat{HS}_{\rm k}(\hat{x}^{-1})$ by appropriately choosing $a$.

This is always possible because the Hilbert series is a meromorphic function of the form

\begin{equation}
HS_{\rm k}(\hat{x})=\frac{\prod_n^N\,(1-\hat{x}^{a_n}\,\hat{y}^{n_n})}{\prod_m^M\,(1-\hat{x}^{c_m}\,\hat{y}^{d_m})}\, ,
\end{equation}
for some $N,\,M$ and some string of exponents $\{a_n,\,b_n,\,c_m,\,d_m\}$. Note that, for simplicity, we have unrefined the G-fugacities (we will come back to this point below). Under $\hat{x}\rightarrow \hat{x}^{-1}$ this goes to

\begin{equation}
HS_{\rm k}(\hat{x}^{-1})=\frac{\prod_n^{N}\,(1-\hat{x}^{-a_n}\,\hat{y}^{n_n})}{\prod_m^M\,(1-\hat{x}^{-c_m}\,\hat{y}^{d_m})}=(-1)^{N-M}\,\prod_{n,\,m}^{N,\,M}\hat{x}^{c_m-a_n}\,\frac{\prod_n^N\,(1-\hat{x}^{a_n}\,(\hat{y}^{-1})^{n_n})}{\prod_m^M\,(1-\hat{x}^{c_m}\,(\hat{y}^{-1})^{d_m})}\, .
\end{equation}
Note that fully unrefining the Hilbert series and expanding around $\hat{x}=1$, the order of the pole is precisely $N-M$. Since the order of the pole coincides with the complex dimension of the instanton moduli space \cite{Benvenuti:2006qr, Benvenuti:2010pq}, which is a hyperk\"ahler variety, we have that $N-M\in 2\,\mathbb{Z}$. Besides, since $HS_{\rm k}$ is invariant under $\hat{y}\leftrightarrow \hat{y}^{-1}$, we have that

\begin{equation}
HS_{\rm k}(\hat{x}^{-1})=\prod_{n,\,m}^{N,\,M}\hat{x}^{c_m-a_n}\,\frac{\prod_n^N\,(1-\hat{x}^{a_n}\,\hat{y}^{n_n})}{\prod_m^M\,(1-\hat{x}^{c_m}\,\hat{y}^{d_m})}=\hat{x}^{2\,a}\,HS_{\rm k}(\hat{x})\, ,
\end{equation}
for some $2\,a\in\mathbb{Z}$.

Given that we can construct the function $\widehat{HS}_{\rm k}(\hat{x})$ which shows the $(\hat{x},\,\hat{y})\rightarrow(\hat{x}^{-1},\,\hat{y}^{-1})$ expected for $SU(2)$ characters, it is then natural to identify $\widehat{HS}_{\rm k}(\hat{x},\,\hat{y},\,\alpha_i)$ with $\mathcal{I}_{\rm inst}^{\rm (k)}$ and $\hat{x}\leftrightarrow x$, $\hat{y}\leftrightarrow{y}$. That is, we conjecture

\begin{equation}
\label{conjecture}
\mathcal{I}_{\rm inst}^{\rm (k)}(x,\,y,\,\alpha_i)\equiv\widehat{HS}_{\rm k}(x,\,y,\,\alpha_i)\, .
\end{equation}

Note that had we explicitly taken the G-fugacities into account by not unrefining when checking the variation of $HS_{\rm k}$ under $\hat{x}\leftrightarrow \hat{x}^{-1}$ nothing would have changed. This is because in the ADHM auxiliary theory multiplets come in real representations --\textit{e.g.} a hyper contains a fundamental and an antifundamental chiral in 4d $\mathcal{N}=1$ notation--. Thus $\alpha_i\leftrightarrow \alpha_i^{-1}$ will also leave $HS_{\rm k}$ invariant and the same manipulation as that done with $\hat{y}$ immediately shows that upon sending $\hat{x}$ to $\hat{x}^{-1}$, $HS_{\rm k}$ only picks an overall factor $\hat{x}^a$.

Finally, note that in the 4d limit where one writes $x=e^{i\,\beta\,(\epsilon_1+\epsilon_2)}$, $y=e^{i\,\beta\,(\epsilon_1-\epsilon_2)}$, $\alpha_i=e^{i\,\beta\,a_i}$ and sends $\beta\rightarrow 0$ the leading behavior is not affected by the $x^a$. Hence, in the 4d limit, the instanton partition function directly coincides with the Hilbert series as shown in \textit{e.g.} \cite{Keller:2011ek}.

\section{Explicit check: pure $SU(2)$ gauge theory and global symmetry enhancement}\label{BO}

We now test our proposal with an explicit computation of a full index and compare it with the known results in the literature. In \cite{Seiberg:1996bd} it was argued that a $USp(2)$ gauge theory with $N_f<8$ fundamental hypers should be a fixed point theory exhibiting, at the origin of the Coulomb branch, an enhanced $E_{N_f+1}$ global symmetry due to massless instantonic particles. This was checked by computing the exact index in  \cite{Kim:2012gu, Iqbal:2012xm}. On the other hand, this theory can be thought of as the $N=1$ case of a $USp(2\,N)$ theory with one antisymmetric hypermultiplet and $N_f$ fundamental hypermultiplets. For $N=1$, when $USp(2)=SU(2)$, the antisymmetric is a singlet and thus decouples. Hence, for all practical purposes, the theory is a pure $SU(2)$ gauge theory with $N_f$ fundamental hypers. On the other hand this theory can be regarded as the worldvolume theory on a stack of $N$ D4 branes probing an $O8^-$ with $N_f$ coinciding D8 branes. In turn, this system can be backreacted finding in the near-brane region an $AdS_6$ geometry \cite{Brandhuber:1999np}, therefore strongly supporting that the dual theory is a fixed point theory.

We will be interested in the $N_f=0$ case for the minimal rank, that is, a pure $USp(2)\equiv SU(2)$ gauge theory, for which we expect an enhanced $SU(2)$ global symmetry due to instantonic particles, and to which our methods are directly applicable. Following the general expressions above, the index will read

\begin{equation}
\label{index}
\mathcal{I}=\int du\,\frac{1-u^2}{u}\,\mathcal{I}_{{\rm pert}}\,\mathcal{I}_{{\rm inst}}\, ,
\end{equation}
where $\int\,du\,\frac{1-u^2}{u}$ is the integration over the gauge group with the $SU(2)$ Haar measure. Furthermore, for the perturbative part it is straightforward to write that

\begin{equation}
\mathcal{I}_{{\rm pert}}={\rm PE}\Big[\,-\frac{x\,(y+y^{-1})}{(1-x\,y)\,(1-x\,y^{-1})}\,\Big((u+u^{-1})^2-1\Big)\,\Big]\, .
\end{equation}

\subsection{The non-perturbative part}

\begin{figure}[h!]
\centering
\includegraphics[scale=.3]{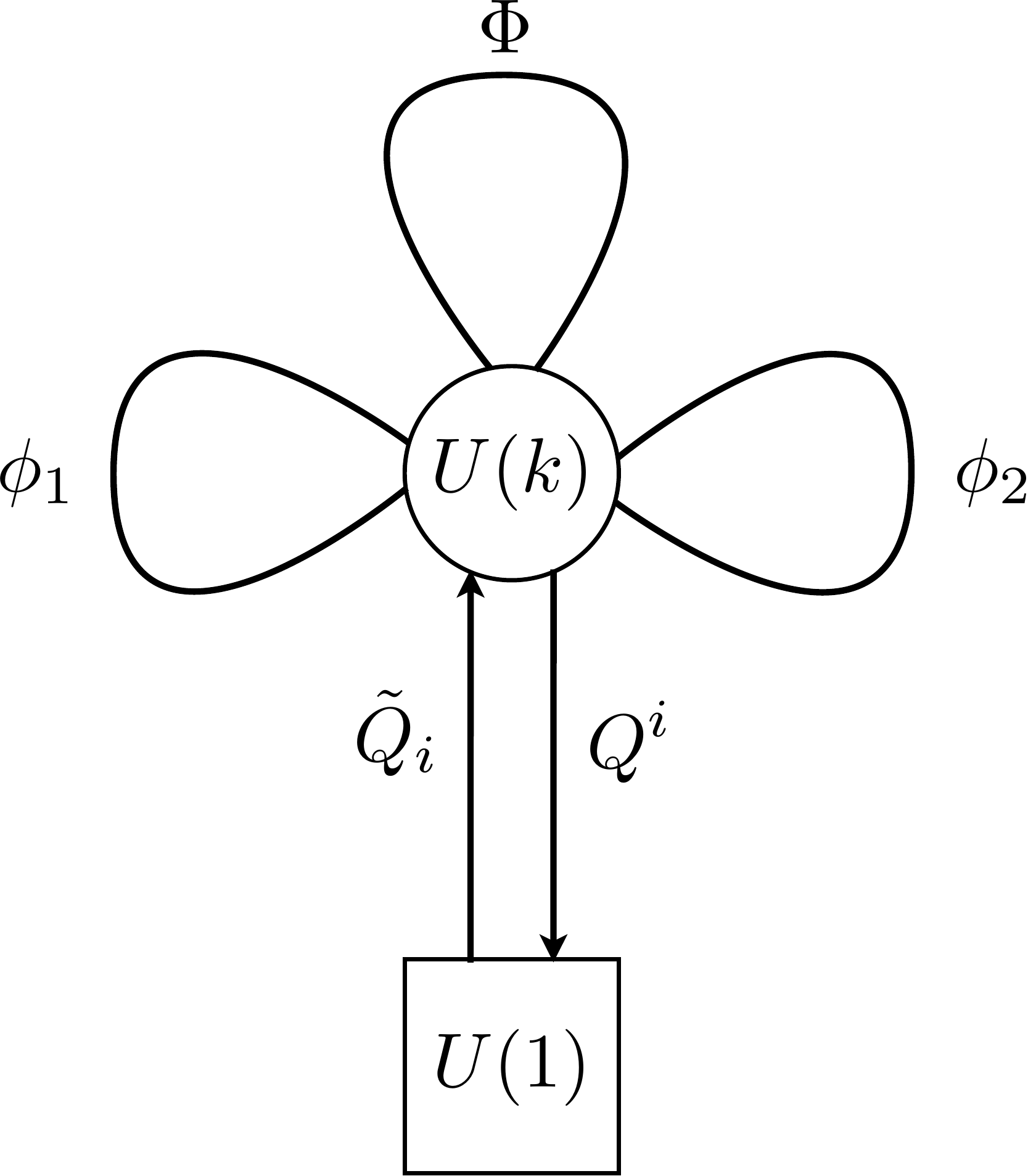}
\caption{Quiver engineering $k$-instantons of $U(2)$ --the same as 1-instanton of $SU(2)\sim USp(2)$--.}
\label{ADHMQuiver}
\end{figure}

Following our conjectured formula, we first need to compute the Hilbert series of the pure $SU(2)$ $k$-instanton on $\mathbb{C}^2$ moduli space. The auxiliary ADHM quiver is shown in figure (\ref{ADHMQuiver}).

The quiver comes with the superpotential

\begin{equation}
W={\rm Tr}\,\Phi\,[\phi_1,\,\phi_2]+\tilde{Q}_i\,\Phi\,Q^i\, .
\end{equation}
Naively there is a global $U(2)$ flavor symmetry. However, the $U(1)$ part coincides with the $U(1)\in U(k)$, and hence the global flavor symmetry will be just $SU(2)$, whose fugacity we will denote by $u$. It will correspond to the $u$ gauge fugacity in eq. (\ref{index}). Besides, there is a global $SU(2)$ acting on $\phi_i$ whose fugacity, as described in section (\ref{motivation}), will be $y$. Finally, there will be an $x$ fugacity for the dimension of the operators. Once the Hilbert series is computed following the methods in \cite{Benvenuti:2010pq, Hanany:2012dm}, as described in section (\ref{motivation}), we will construct the associated function invariant under $x\leftrightarrow x^{-1}$.

\subsubsection{One-instanton}

This corresponds to the case $k=1$ above. The $\mathcal{F}$ flat part of the Higgs branch yields

\begin{equation}
g_{\mathcal{F}^{\flat}}^{(1)}={\rm PE}\Big[ (y+y^{-1})\,x+ (u+u^{-1})\,(r+r^{-1})\,x-x^2\Big]\, ,
\end{equation}
where $r$ is the ADHM $U(1)$ gauge group fugacity. Upon integrating over it we find the one-instanton Hilbert series $HS_{\rm 1}$

\begin{equation}
HS_{\rm 1}=\frac{ \left(1+x^2\right)}{\left(1-\frac{x^2}{u^2}\right) \left(1-u^2 x^2\right) \left(1-\frac{x}{y}\right) (1-x y)}\, .
\end{equation}
As expected, this is not invariant under $x\leftarrow x^{-1}$. Imposing such invariance by multiplying by the adequate power of $x$ fixes $\mathcal{I}_{\rm inst}^{\rm{(1)}}=x^2\,HS_{\rm 1}$, which explicitly reads

\begin{equation}
\mathcal{I}_{\rm inst}^{\rm{(1)}}=\frac{x^2 \left(1+x^2\right)}{\left(1-\frac{x^2}{u^2}\right) \left(1-u^2 x^2\right) \left(1-\frac{x}{y}\right) (1-x y)}\, .
\end{equation}
One can compare this expression with previous results in the literature (see \textit{e.g.} \cite{Kim:2012gu}) obtaining exact agreement.

\subsubsection{Two instantons}

This corresponds to the case $k=2$ in fig. (\ref{ADHMQuiver}). In this case, since $N_c=N_f$ the gauge group is not entirely broken. We can however compute the 2-instanton moduli space Hilbert series by brute force using \verb+Macaulay2+ \cite{Mac2}. One can check that, enforcing the symmetry $x\leftrightarrow x^{-1}$, one finds $\mathcal{I}_{\rm inst}^{\rm (k)}=x^4\,HS_{\rm 4}$. The final result is a bit cumbersome and can be written as

\begin{equation}
\mathcal{I}_{{\rm inst}}^{\rm (2)}=\frac{\mathcal{P}}{\mathcal{Q}}\, ,
\end{equation}
where

 \begin{eqnarray}
 \mathcal{P}&=& -u^6\, x^7\, y^4+u^4\, x^9\, y^4+u^8\, x^9\, y^4+u^4\, x^{11} \,y^4+u^8\, x^{11}\, y^4+u^4\, x^{13}\, y^4+u^8\, x^{13}\, y^4-u^6 \,x^{15}\, y^4-u^6\, x^4\, y^5 \nonumber \\ \nonumber && -2\, u^6\, x^6\, y^5-2\, u^6\, x^8\, y^5+u^4\, x^{10}\, y^5-u^6\, x^{10}\, y^5+u^8\, x^{10}\, y^5 +u^4\, x^{12}\, y^5-u^6\, x^{12}\, y^5+u^8\, x^{12}\, y^5\nonumber \\ \nonumber && -2\, u^6\, x^{14}\, y^5-2\, u^6\, x^{16}\, y^5-u^6\, x^{18}\, y^5-u^6\, x^7\, y^6+u^4\, x^9\, y^6+u^8\, x^9\, y^6 +u^4\, x^{11}\, y^6+u^8\, x^{11}\, y^6\nonumber \\ \nonumber && +u^4\, x^{13}\, y^6\,+u^8\, x^{13}\, y^6-u^6\, x^{15}\, y^6\, ,
\end{eqnarray}

and

 \begin{eqnarray}
 \mathcal{Q}&=&\left(x^2+u^4\, x^2-u^2 \left(1+x^4\right)\right) \left(x+y+x^2\, y+x\, y^2\right) \left(y+x^2 \,y-x \left(1+y^2\right)\right)^2\nonumber \\ \nonumber && \left(u^4\, y+x^6\, y-u^2\, x^3 \left(1+y^2\right)\right) \left(y+u^4\, x^6 \,y-u^2\, x^3 \left(1+y^2\right)\right)\, .
\end{eqnarray}

To give a flavor of the result, let us quote the fully unrefined index

\begin{equation}
\mathcal{I}_{{\rm inst}}^{\rm (2)}=\frac{x^4 \left(1+x+3 x^2+6 x^3+8 x^4+6 x^5+8 x^6+6 x^7+3 x^8+x^9+x^{10}\right)}{(1-x)^8 (1+x)^4 \left(1+x+x^2\right)^3}\, .
\end{equation}
As in the one-instanton case, one can explicitly compare these expressions with results in the literature (see \textit{e.g.} \cite{Kim:2012gu}) obtaining again exact agreement.

Note that, up to the $x^4$ factor enforcing the $x\leftrightarrow x^{-1}$ invariance, the result is a palindrome as expected for a hyperk\"ahler moduli space. Besides, the order of the pole at $x=1$ is 8. Since  the geometric Hilbert series of $\mathbb{C}^2$ has a pole at $x=1$ of order 2, this means that the reduced instanton moduli space for 2 $SU(2)$ instantons is 6 complex dimensional, in agreement with \cite{Hanany:2012dm}.

\subsection{The full index}

As just shown, the instanton partition functions computed using our technique exactly agree with the expected results in the literature. For the sake of completeness, let us now compute the full index just by combining the results above as described in section (\ref{motivation}) and integrating over $u$ in (\ref{index}). Note that the $k$-instanton index enters at order $x^{2\,k}$. Hence up to the 2-instanton order computed here we can at most go up to $x^5$. Up to that order we find

\begin{eqnarray}
\mathcal{I}&=&1+\left(1+\frac{1}{q}+q\right) x^2+\frac{(1+q)^2 \left(1+y^2\right)}{q y}\,x^3\\ \nonumber && +\frac{\left(y^2+q^4 y^2+q \left(1+y^2\right)^2+2 q^2 \left(1+y^2\right)^2+q^3 \left(1+y^2\right)^2\right)}{q^2 y^2}\,x^4 \\ \nonumber && +\frac{\left(1+y^2\right) \left(y^2+q^4 y^2+q \left(1+y^2\right)^2+q^3 \left(1+y^2\right)^2+q^2 \left(2+3 y^2+2 y^4\right)\right)}{q^2 y^3}\,x^5\\ \nonumber && +\mathcal{O}(x^6)\, .
\end{eqnarray}
It is a straightforward exercise to show that this expression precisely agrees with eq. (4.9) in \cite{Kim:2012gu}. In particular, one can see the appearance of the $SU(2)$ characters in $q$, hence explicitly showing the enhanced global $SU(2)$ symmetry.

All in all, we have shown the explicit agreement between the index computed following our prescription with the known results in the literature up to order $x^6$. Going to arbitrarily higher orders is a tedius but straightforward exercise. Note in particular that the prescription to select the poles contributing is completely fixed: just those with positive power of $x$ --this is inherited from the original $\mathbb{R}$ nature of $x$ in $HS$--.

\section{Exact index for a pure $U(1)$ 5d SCFT} \label{U(1)}

Let us now apply our technique to the case of pure $U(1)$ gauge theory. Although $U(1)$ instantons are singular, we can consider the non-commutative deformation removing the small instanton singularity. The exact index for a pure $U(1)$ SCFT is

\begin{equation}
\mathcal{I}=\int\,\frac{du}{u}\,\mathcal{I}_{\rm inst}\,\mathcal{I}_{\rm pert}\, .
\end{equation}
Here, the perturbative part is simply

\begin{equation}
\mathcal{I}_{{\rm pert}}={\rm PE}\Big[\,-\frac{x\,(y+y^{-1})}{(1-x\,y)\,(1-x\,y^{-1})}\,\Big]\, .
\end{equation}
Note that it does not depend on the $U(1)$ fugacity $u$.

\begin{figure}[h!]
\centering
\includegraphics[scale=.3]{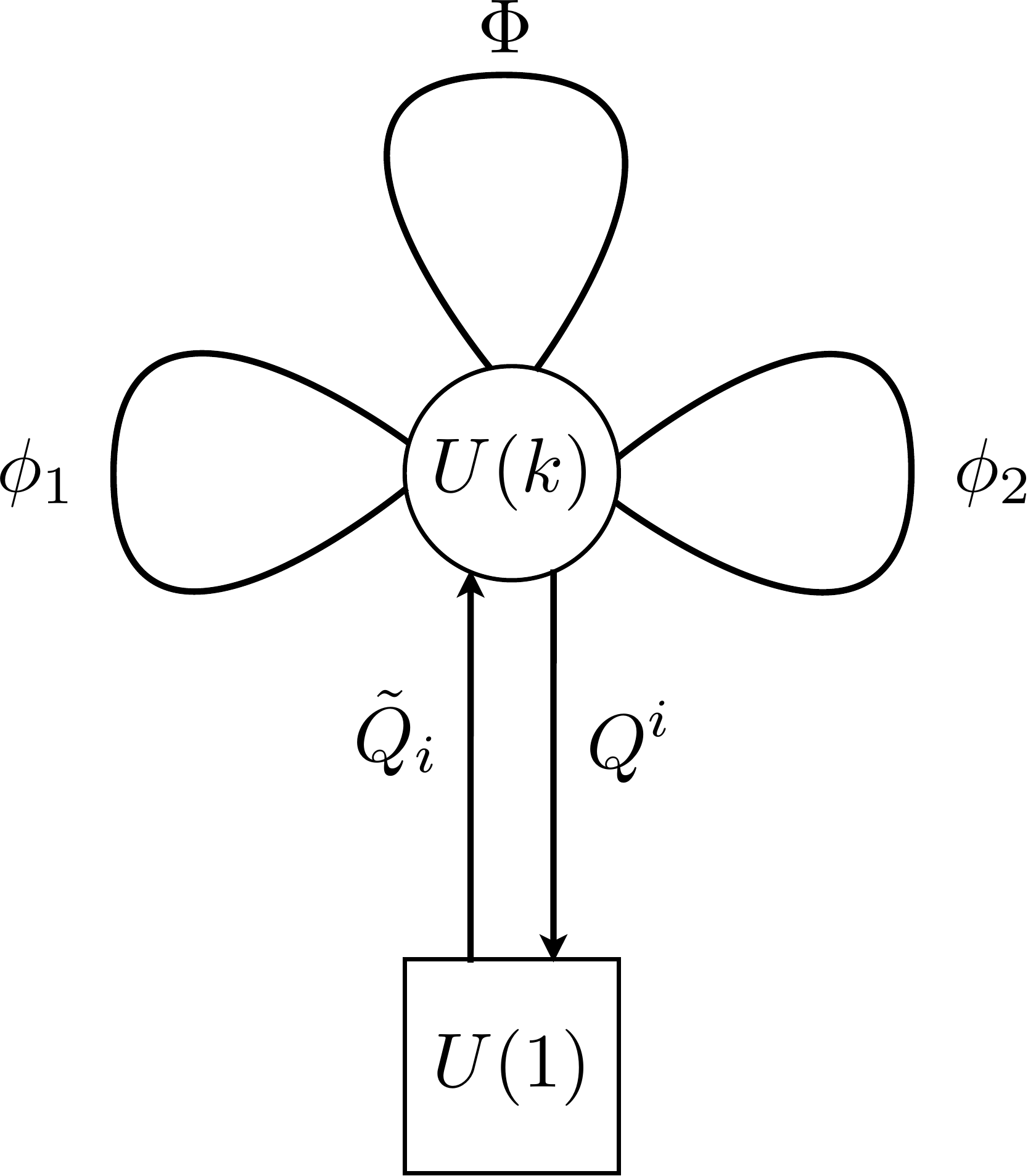}
\caption{Quiver engineering $k$-instantons of $U(1)$.}
\label{ADHMQuiverU(1)}
\end{figure}

As for the instanton contribution, following our recipe, they arise from the Hilbert series on the Higgs branch of the auxiliary ADHM quiver depicted in fig.(\ref{ADHMQuiverU(1)})

Denoting by $r$ the ADHM $U(1)$ fugacity, the $F$-flat part of the Higgs branch yields

\begin{equation}
g_{\mathcal{F}^{\flat}}^{(1)}={\rm PE}[\,(y+y^{-1})\,x+x\,(r+r^{-1})-x^2\,]\, .
\end{equation}
Thus, upon integrating over the ADHM auxiliary $U(1)$ gauge group and imposing the $x\leftrightarrow x^{-1}$, the 1-instanton index  is easily seen to be (see also \cite{Poghossian:2008ge})

\begin{equation}
\mathcal{I}^{\rm (1)}_{\rm inst}=\frac{x}{(1-x\,y)\,(1-x\,y^{-1})}
\end{equation}
We can compute higher instanton indices. However, these are technically slightly more involved, as the auxiliary AHDM gauge group will not be completely higgsed for $k>1$. Computing by brute force the Hilbert series using  \verb+Macaulay2+ \cite{Mac2}, one can see that

\begin{equation}
\mathcal{I}^{\rm (2)}_{\rm inst}=\frac{x^2\,(1+x^2)}{(1-x\,y)\,(1-x\,y^{-1})\,(1-x^2\,y^2)\,(1-x^2\,y^{-2})\,}=\frac{\mathcal{I}(\vec{x})^2+\mathcal{I}(\vec{x}^2)}{2}
\end{equation}
This is just the second coefficient of the Taylor expansion in $q$ of

\begin{equation}
\label{InstU(1)}
\mathcal{I}^{\rm N}_{\rm inst}(q)={\rm PE}\Big[\,\frac{x\,q}{(1-x\,y)\,(1-x\,y^{-1})}\,\Big]\, .
\end{equation}
In fact, eq.(\ref{InstU(1)}) is to be expected. Since we have a pure $U(1)$ theory, instantons are non-interacting pointlike particles, and hence the $k$-instanton contribution is inherited from the 1-instanton. Furthermore, the fugacity $q$ above counts instanton number, and hence indeed coincides with the original instanton fugacity. Indeed, one can check that higher order coefficients of this expansion agree with the instanton index computed from the appropriate $k$ ADHM quiver.

Taking into account the south pole contribution --which is just the same upon doing $q\rightarrow 1/q$-- we find the instanton index

\begin{equation}
\mathcal{I}_{\rm inst}={\rm PE}\Big[\frac{x}{(1-x\,y)\,(1-x\,y^{-1})}\,(q+q^{-1})]\,\Big]\, .
\end{equation}
Note that this coincides with the contribution to the index of a hypermultiplet, $q$ playing the role of the global symmetry fugacity. Hence, all in all, the exact index for the pure $U(1)$ theory is

\begin{equation}
\mathcal{I}={\rm PE}\Big[\,-\frac{x\,(y+y^{-1})}{(1-x\,y)\,(1-x\,y^{-1})}+\frac{x\,(q+q^{-1})}{(1-x\,y)\,(1-x\,y^{-1})}\,\Big]\, .
\end{equation}

\section{Conclusions} \label{conclusions}

In this note we have made explicit the connection between the Hilbert series of the moduli space of instantons with the instanton index of 5d pure gauge theories. Since the Nekrasov instanton partition formula involves compact symmetries, while the Hilbert series is a generating function depending on an $\mathbb{R}$ fugacity, the mapping between Hilbert series and instanton index requires ``covariantizing" the Hilbert series so that the fugacity counting the dimension of operators is converted into an $SU(2)$ fugacity. We used the exact index computation of the pure $USp(2)\sim SU(2)$ theory  to show the agreement of our proposal, displaying explicit computations up to 2 instantons. It should be stressed that, although we are not displaying them to keep the presentation contained, similar consistency checks have been performed up to higher $k$ as as well as on diverse other instanton indices for higher $SU(N)$ groups obtaining perfect agreement with the results in the literature.

It is immediate to ask how our procedure can be extended to compute instanton indices of 5d theories with extra matter in arbitrary representations.  While it is not clear how to extend the computation of the ``Hilbert series" to moduli spaces of flavored instantons, comparison with \cite{Kim:2012gu} suggests that one can incorporate the flavor contribution simply by  multiplying $g_{\mathcal{F}^{\flat}}$ by an extra factor incorporating information about the extra matter.

It would be very interesting to apply these techniques --or else the more standard methods well-known in the literature-- to the computation of indices for the quiver theories introduced in   \cite{Bergman:2012kr,Bergman:2012qh}, in particular clarifying wether enhanced symmetries do indeed arise in the quiver case. As these theories do admit an $AdS_6$ dual, it would also be very interesting to study the large $N$ version of the index and compare with the SUGRA dual. It is natural to expect that instantons correspond to spinning D0 branes, and it would be very interesting to compute the large $N$ instanton index from the gravity dual.

\section*{Acknowledgements}	
We are grateful to Noppadol Mekareeya for very useful conversations. We would also like to thank Oren Bergman for collaboration at early stages of this project as well as for very useful conversations. G.Z is supported in part by the Israel Science Foundation under grant no. 392/09, and the US-Israel Binational Science Foundation under grant no. 2008-072. D. R G is partially supported by the research grants MICINN-09-FPA2009-07122 and the Ramon y Cajal Fellowsip MEC-DGI-CSD2007-00042.

\end{document}